# Oxygen isotope anomalies of the Sun and the original environment of the Solar system

Jeong-Eun Lee[1, 2], Edwin A. Bergin[3], and James R. Lyons[4, 5]

*[1]Department of Astronomy and Space Science, Astrophysical Research Center for the Structure and Evolution of the Cosmos, Sejong University, Seoul 143-747, Korea, jelee@sejong.ac.kr*

*[2]Physics and Astronomy Department, The University of California at Los Angeles, PAB, Box 951547, Los Angeles, CA 90095-1547 (Hubble Fellow)*

*[3]Department of Astronomy, The University of Michigan, 500 Church Street, Ann Arbor, Michigan 48109-10424*

*[4]Institute of Geophysics and Planetary Physics, [5]Department of Earth and Space Sciences, University of California, Los Angeles, California 90095, USA*

**Abstract**

**We present results from a model of oxygen isotopic anomaly production through selective photodissociation of CO within the collapsing proto-Solar cloud. Our model produces a proto-Sun with a wide range of $\Delta^{17}O$ values depending on the intensity of the ultraviolet radiation field. Dramatically different results from two recent Solar wind oxygen isotope measurements indicate that a variety of compositions remain possible for the solar oxygen isotope composition. However, constrained by other measurements from comets and meteorites, our models imply the birth of the Sun in a stellar cluster with an enhanced radiation field and are therefore consistent with a supernova source for $^{60}Fe$ in meteorites.**

## I. Introduction

The oxygen isotope composition of the Sun is central to understanding the oxygen isotope evolution of the Solar System as recorded in meteorites. Recent theories for oxygen isotopes in the solar nebula (Clayton 2002; Yurimoto et al. 2004; Lyons et al. 2005) suggest a linkage between the bulk Sun, calcium-aluminum inclusions (CAIs) in meteorites, and the ultraviolet radiation field in the vicinity of the proto-solar nebula, and, if correct, place strong constraints on the formation environment of the Solar system. In these models isotope-selective photodissociation of CO, either in the solar nebula or parent cloud, results in a mass-independent isotope signature in atomic oxygen that is incorporated into nebula $H_2O$. Segregation of water ice to the nebular midplane imparts a positive mass-independent signature to the planets, but was assumed to not alter the isotopic signature of the bulk proto-Sun. We show here that this assumption is in general not correct.

Measurements of oxygen isotopes in lunar metal grains (Hashizume et al. 2005; Ireland et al. 2006), are potential proxies for the isotopic composition of the convective zone of the Sun. Earlier work on N isotopes in lunar regolith grains demonstrated the presence of both solar and non-solar N components with very distinct $\delta^{15}N$ values (Hashizume et al. 2000). The O isotope results from Hashizume and Chaussidon (2005) and Ireland et al. (2006) are quite disparate, but in view of the N results this is not entirely unexpected. Expressed as the mass-independent component of the isotopic delta values, $\Delta^{17}O_{SMOW} = \delta^{17}O_{SMOW} - 0.52\delta^{18}O_{SMOW}$, which eliminates the mass-dependent fractionations affecting Solar wind ions, the measured values are down to $\sim$ -20 ‰ in one case (Hashizume et al. 2005) and +26 (±3) ‰ in the other (Ireland et al. 2006). Clayton (2002) predicted that the Sun would have $\Delta^{17}O \sim$ -20 ‰ based on the isotopically-lightest CAIs, which were assumed to have oxygen isotope ratios similar to the bulk parent cloud. This assumption is open to debate. According to Wiens et al. (2004), the Solar values measured from the Solar wind and the photosphere are not different from values measured from meteorites and comets at the 2-σ uncertainty level. We may argue that a planetary oxygen isotope composition of $\sim$ 0 ‰ is a possible solar composition as well. Here we explore the possibility that CO self-shielding can also explain various values of Solar $\Delta^{17}O$.

Irradiated disk models (Lyons et al. 2005) likely do not provide enough mass to alter bulk Solar oxygen isotope ratios, except possibly by a late addition of water–rich material to the Solar convective zone. For a convective zone mass of 0.03 $M_\odot$, and assuming cometary material with water with $\Delta^{17}O \sim$ +100 ‰, about 50 Earth masses of comets (a mass comparable to the Oort cloud) are needed to increase $\Delta^{17}O$ of the convection zone by $\sim$ 20 ‰. Such large external input of planetary material to stellar convective zones does not in general appear to be supported by recent observations of extra-Solar planetary systems (Ecuvillon et al. 2006). By comparing stellar atmosphere elemental abundances versus elemental condensation temperature for stars with and without planets, Ecuvillon et al. (2006) concluded that the metallic excess observed in planet host stars is mainly 'primordial', the result of galactic chemical evolution rather than 'pollution' by planets. Here we present results of a model that examines the question of whether material that is isotopically enriched within the proto-solar cloud that collapses to form the star and planet-forming disk can impart these anomalies directly to the proto-Sun.

## II. A model of a proto-solar molecular core collapse

Fig. 1 describes schematically the sequence of star formation. Stars form by the gravitational collapse of a slightly rotating dense molecular cloud core with the accretion through a disk, which forms due to the intrinsic rotation in the parent molecular cloud. This figure also describes the development of oxygen isotopic anomalies in the molecular core and the transfer of the anomalies to the disk from the envelope. The envelope of the dense molecular core is reasonably approximated with a symmetric sphere, making a self-consistent calculation of the coupled physical and chemical evolution practicable. We use a chemo-dynamical model (Lee et al. 2004) that computes simultaneously the inward transfer and chemistry of gas and dust within a collapsing molecular core (i.e. prior to the formation of solar nebula). The results at the inner radius of the molecular core model can be considered as initial conditions to the outer disk. We do not model the processes in the disk, instead we use a simple calculation to show the possible variation of Solar values, as described in Section IV.

The model of a collapsing molecular core describes the evolutionary stages when accretion is still active in the star formation process and assumes that a central concentrated molecular core collapses from "inside-out" (Shu 1977). In the inside-out collapse model, we use a 10 K isothermal sphere with an outer radius of 30,000 AU and an effective sound speed of 0.222 km s$^{-1}$. The collapse model has been combined with dust continuum radiative transfer and gas energetics to calculate dust and gas temperatures self-consistently (see Lee et al. 2004 for more details). The model core is assumed to evolve through a sequence of 7 Bonnor-Ebert spheres (Ebert 1955, Bonnor 1956), which have flat densities at small radii and power-law density profiles at large radii and are believed to describe density structures of dense molecular cores prior to collapse (Evans et al. 2001), for $5 \times 10^5$ years before collapse. Fig. 2 describes the evolution of a dense molecular core before and after collapse. The central density grows with time before collapse, and collapse occurs from the center and moves outward (inside-out collapse). Therefore, the central density decreases with time after collapse.

The chemical evolution was solved using the UMIST kinetics database (Millar et al. 1997) that has been previously adopted to include interactions of molecules with grains and a full isotopic chemistry (Bergin et al. 1997a; Bergin et al. 2006). In order to include the isotope-selective dissociation of CO, the self-shielding functions of van Dishoeck and Black (1988) (Table 5) were adopted (see the next section).

In our chemical model, the molecular core is assumed to have initially the following composition (with respect to H nuclei): X(He)=0.08, X(CO)=$1.4 \times 10^{-4}$, X(H$_2$O)=$0.6 \times 10^{-4}$, X(Fe)=$3 \times 10^{-8}$, and X(Fe+)=$6 \times 10^{-12}$. The isotopic ratios of $^{16}$O/$^{18}$O and $^{16}$O/$^{17}$O are assumed to be 500 and $5.16 \times 500$, respectively, for the initial abundances of C$^{18}$O, C$^{17}$O, H$_2$$^{18}$O, and H$_2$$^{17}$O. We assume that initially, all CO is in gas while all H$_2$O is in ice. Excluding water ice (which is subject to hydrogen bonding), binding energies to bare silicate grains are used throughout. The binding energies (E/k, where k is the Boltzmann constant) for CO and H$_2$O are 1180 and 4800 K, respectively.

In this model, we include photodesorption as well as thermal desorption of ice. The photodesorption yield of other molecules except for H$_2$O and CO is set to be $10^{-4}$ (Greenberg 1973). We adopt $10^{-3}$ for the photodesorption yield of H$_2$O (Westley et al. 1995) and CO (Oberg et al. 2007). Photodesorption can affect the oxygen isotopic anomalies in ice at the edge of the proto-solar cloud. Photodesorption of $H_2O$ results in the (mass-dependent) photodissociation of $H_2O$, which in turn, provides oxygen for CO formation and subsequent photodissociation, the key process for mass-independent isotopic fractionation of oxygen.

Finally, chemical evolution has been calculated by following each gas parcel, whose physical conditions at each time step can be traced based on the continuity equation in Lagrangian coordinates (see Equations 1, 2, and 3 of Lee et al. 2004). Once the chemical calculation is done for all gas parcels, they are transferred to an Eulerian coordinate system to describe the distribution of abundances at any given time. Therefore, this chemical model combined with a dynamical model can describe the transfer of the oxygen isotopic anomalies, which are imprinted in infalling material, inward.

### III. Isotopic Selective Photodissociation

The creation of anomalies can occur by the selective dissociation of $C^{16}O$ and its isotopologues ($C^{18}O$ and $C^{17}O$) because the dissociation rate sensitively depends on the abundance of each isotopologue. In particular, CO dissociation from external radiation occurs via a line process and CO molecules closer to radiation source can effectively shield other CO molecules by absorbing radiation. Due to the dependence of the self-shielding on column densities, $C^{18}O$ and $C^{17}O$, which are less abundant than $C^{16}O$ by factors of 500 and $5.16 \times 500$, respectively, are dissociated in deeper regions. Thus, there will exist a layer of atomic $^{18}O$ (and $^{17}O$) when all gaseous $^{16}O$ is locked in CO. This process is well documented in interstellar molecular clouds (van Dishoeck et al. 1988). The free $^{18}O$ (and $^{17}O$) can freeze onto grains and react with hydrogen to make isotopic enriched water ice. Thus, there will exist a water ice layer with an enrichment of $^{18}O$ and $^{17}O$ relative to $^{16}O$ (above the values estimated for the interstellar medium).

Similar to the previous examination of isotopic anomalies in the solar nebula (Lyons et al. 2005), we quantify the CO isotopologue photodissociation rates and self-shielding using the parameterization developed for molecular cores (van Dishoeck et al. 1988). All photodissociation rates are scaled by the enhancement factor, $G_0$, which is a measure of the strength of the local far-ultraviolet (FUV) radiation field relative to the standard interstellar radiation field (Habing 1968). The self-shielding becomes more and less effective with density and $G_0$, respectively. As described in Fig. 2, for a given $G_0$, the FUV photons penetrate less deeply as the density grows before collapse since FUV attenuation by grain absorption increases at a given radius (i.e. the total column/optical depth increases). A scale to measure the amount of shielding of the radiation field by grain absorption at optical wavelengths is the visual extinction ($A_V$), and the photodissociation rate is proportional to $G_0 \times \exp(-3 \times A_V)$. Fig. 3 shows the distribution of $A_V$ for 7 Bonnor-Ebert spheres, that is, at the time steps of $5 \times 10^5$, $2.5 \times 10^5$, $1.25 \times 10^5$, $6.5 \times 10^4$, $3.5 \times 10^4$, $1.5 \times 10^4$, and $5 \times 10^3$ years before collapse.

We investigate various FUV enhancement factors from the normal interstellar radiation field ($G_0$=1) to an extremely enhanced FUV field such as $G_0$=$10^5$, which is representative of a massive O star in the near vicinity (0.1 – 0.3 pc) of the proto-Solar cloud as it was collapsing. There is some evidence for the latter case in the literature from the inference that $^{60}$Fe was present in the solar nebula during early evolutionary stages (Shukolyukov and Lugwair 1993; Tachibana and Huss 2003). The presence of $^{60}$Fe has been suggested as primary evidence that the Sun formed near a massive O star (see Wadhwa et al. 2007). $^{60}$Fe is believed to be produced during the core collapse of a supernova and externally seeded to the disk (Ouellette et al. 2005) or into the proto-solar molecular cloud/core (Cameron et al. 1977; Gounelle 2006). There is a caveat in our 1-dimensional model for the FUV field enhanced by a nearby massive star since the irradiation by the nearby star is not spherically symmetric. However, the material accreted from the envelope to the disk will mix and distribute evenly in the azimuthal direction in the disk although the effect of the selective CO photodissociation obtained in the envelope is reduced by a factor of 2~4 by the mixing.

Fig. 4 and 5 compare results from various models with different values of $G_0$ at 4 time steps before collapse and two time steps after collapse. $4.6 \times 10^5$ years here represents the timescale when the proto-Sun achieves 1 $M_\odot$ through accretion. The $^{18}$O anomaly is calculated from water ice, and it is relative to the standard mean ocean water (SMOW), $\delta^{18}O_{SMOW}$ from the equation, $\delta^{18}O_{SMOW}$=$\delta^{18}O_{MC}$−50 and $\delta^{18}O_{MC}$=$10^3 \times$ ([H$_2$$^{18}$O$_{ice}$]/[H$_2$O$_{ice}$]/(1/500)−1) where $\delta^{18}O_{MC}$ is the anomaly relative to interstellar molecular clouds. We assume the initial condition of $\delta^{18}O_{SMOW}$ in the molecular cloud is −50 ‰ according to Yurimoto and Kuramoto (2004). It could also be argued that we should use the −80 ‰ as the initial condition based on the isotopically lightest chondrule measured (Kobayashi et al. 2003). Isotopic anomalies are reached in the steady state pre-collapse phase. Before collapse, in most cases, the largest anomalies are observed toward the cloud edge. However, at early evolutionary stages before collapse, due to the coupled interaction between CO self-shielding and UV attenuation by grain absorption, there is a variable level of anomaly imprinted at the core center. This reaches maximum at $G_0$=25, where the isotope-selective photodissociation implants the highest levels of anomalies in the water ice in the core center. $\delta^{18}O_{SMOW}$ developed in water ices at small radii at earlier time steps will be preserved thoughout the evolution of the proto-solar molecular core before collapse. This owes to the fact that the dust temperatures are below the water ice evaporation temperature; in addition, as the volume and column density grows FUV photons cannot penetrate to the deep center to photodissociate more C$^{18}$O and C$^{17}$O, or to photodesorb water. Thus, once collapse begins, the anomalies built at large radii move inward with infalling material, as seen at the time steps of $2 \times 10^5$ and $4.6 \times 10^5$ years after collapse in Fig. 4 and 5.

In the following we illustrate how anomalies are imprinted in ices due to isotope selective photodissociation. In Fig. 6 we show the distribution of the abundances of CO and O isotopes as well as $\delta^{18}O_{SMOW}$ at $2.5 \times 10^5$ years before collapse with respect to $A_V$, at



$G_0$=1. In this model $C^{16}O$ is photodissociated at $A_V < \sim 1$ mag while $C^{18}O$ and $C^{17}O$ are photodissociated at $A_V < \sim 2$ mag when exposed to the normal interstellar radiation field. As a result, the selective photodissociation occurs at $A_V < \sim 2$ mag, producing a peak at $A_V = 1$. In our model, all isotopes are completely shielded at $A_V > \sim 2$ mag (log r < 3.3). However, for $1 < A_V < 2$ mag ($3.3 < \log r < 4$), $C^{16}O$ is almost completely self-shielded while $C^{18}O$ and $C^{17}O$ are partly self-shielded with their photodissociation rates decreasing inward leading $\delta^{18}O_{SMOW}$ to decrease inward. In contrast, at $A_V < 1$ (log r > 4), $\delta^{18}O_{SMOW}$ decreases outward since almost all $C^{18}O$ is photodissociated while $C^{16}O$ is partly photodissocated with its photodissociation rate increasing outward.

As described, in the isotope-selective photodissocation, there are four regimes: (1) completely shielded for both $C^{16}O$ and $C^{18}O$, (2) completely shielded for $C^{16}O$ but partly shielded for $C^{18}O$ with the $C^{18}O$ photodissociation rate increasing outward, (3) completely photodissociated in $C^{18}O$ but partly photodissociated in $C^{16}O$ with the $C^{16}O$ photodissociation rate decreasing inward, and (4) completely photodissociated in both $C^{16}O$ and $C^{18}O$. As a result, the anomaly peak is built between regimes (2) and (3). Fig. 7a presents how the anomaly peak shifts with changing $G_0$. $\delta^{18}O_{SMOW}$ for $G_0$=10, 25, and 100 peaks at $A_V$=1.6, 1.85, and >2.2 mag, respectively. This shift is caused by the photodissociation rates that are a function of $G_0$ and $A_V$ which, excluding self-shielding, scales as $G_0 \exp(-3 \times A_V)$ $(1-\exp(-1.5 \times A_V))/A_V$ (van Dishoeck and Black 1988). In Fig. 7(b), we show the $C^{16}O$ photodissociation rates for variable radiation field and also, $k_{max}$, the photodissociation rate where $\delta^{18}O_{SMOW}$ reaches its maximum in the case of $G_0$=1 (at $A_V$=1 mag). As can be seen the depth at which the photorate decay reaches $k_{max}$ is the same depth where the maximum anomaly is found in Fig. 7a ($A_V$=1.6, 1.85, and ~2.3 mag for $G_0$=10, 25, and 100, respectively).

In some cases (e.g. $G_0$=25) we find high values for $\delta^{18}O_{SMOW}$ even in the core center. The previous discussion illustrates how anomalies are imprinted at the core edges (and then carried inward by collapse). However, it does not explain how, in a few cases, the heavy isotopes are enriched in the dense core center. Here the answer lies in the evolution of extinction shown in Fig. 3. At early times the cloud is more diffuse and radiation can penetrate throughout and as the core contracts the column and volume density rises providing increased radiation shielding. At the earliest time, a combination of radiation field strength and total $A_V$ can lead to cases where anomalies are imprinted throughout the diffuse cloud, essentially as a starting condition for collapse. This is the case for $G_0$=25, 100, and at later times $G_0$=$10^5$.

To illustrate this issue, in Fig. 8 we show the ratio of photodissociation rates for two models, $G_0$=25, with high anomalies at all radii during the initial evolutionary stage, and $G_0$=100, which only shows anomalies increasing towards the core center (see Fig. 4). In Fig. 8 the ratio is flat for r < 3000 AU at $G_0$=25, but at similar radii is increasing toward the center when $G_0$=100. This somewhat mirrors the structure seen for $\delta^{18}O_{SMOW}$ in Fig. 4 in the sense that large and near constant anomalies are seen when $G_0$=25; in contrast



anomalies are rising towards the center for $G_0$=100. Thus in these models a photo-rate ratio of ~6-10 is the most efficient at leading to heavy oxygen isotope enrichment. Lower ratios are where both isotopologues are either fully shielded or dissociated with no subsequent anomaly.

The above discussion places focus on the results for the first time step where $G_0$=25 and 100 produce isotopic enrichments deep in the core. Other models do not produce anomalies at this stage because, in the case of $G_0$=1, the center is fully shielded, while for $G_0 > 100$ both species are dissociated even in the center. However, at the second step the extinction rises and the relation between CO self-shielding and UV attenuation by grain absorption is altered. Here, selective photodissociation at $G_0$=$10^3$, $10^4$, and $10^5$ have similar effectiveness as $G_0$=1 - $10^2$ in the first time step. However, CO self-shielding turns on at smaller radii, which, in turn, makes the anomaly rise appear at smaller radii than at lower $G_0$.

*The dependence of isotopic anomalies on $G_0$ at small radii before collapse is significant because it suggests much of the cloud mass can be implanted with anomalies. Thus, the Solar oxygen isotope ratio can be affected by the original environment (i.e., the collapsing gas cloud to form the Sun).* Once collapse begins, material with imprints of high anomalies at large radii move inward, and the enrichment of heavier isotopes become larger since remaining mass in the cloud envelope decreases with collapse, thus more FUV photons can penetrate to dissociate more CO isotopes.

## IV. Material Fractionation between Water Ice and CO Gas

There are two separate processes related to the enhancement of heavier oxygen isotopes in our Solar System. First we must have an enhancement of the heavier oxygen isotopes in ice and their deficiency in gas. Second, solid bodies in the inner solar system condensed out of evaporated solids and mixed with the gas (and the Sun formed from this mixed material). Therefore there must be a mechanism to segregate ice and gas components to lead to anomalies in the forming bodies. In the following we will refer to this mechanism as material fractionation.

Our model of molecular core collapse shows that the first process can be operative during the formation and collapse of the proto-solar cloud. A number of recent authors have also suggested that material fractionation in the inner nebula is also possible (Cuzzi et al. 2004; Ciesla et al. 2006). In their models the ice coated grains accreted to the outer disk at the early stages will settle in the midplane of the disk, grow in size, and move inward to small radii due to gas drag. In contrast, the gas can be removed from the disk midplane through advection, diffusion, or dispersal (Yurimoto et al. 2007). Cuzzi et al. (2004) have shown that the drifting grains coated with icy mantles containing the enhancement in heavier oxygen isotopes evaporate when crossing the water evaporation front (snow line). As a result, the inner disk is enhanced in $H_2^{17}O$ and $H_2^{18}O$ vapours, which cool down,



condense again, and finally constitute planets and meteorites in later evolutionary stages. It is known from studies of other systems that disks exist from early stages in the star formation process and accretion is mediated by the disk (Terebey et al. 1984; Hartmann 1998). In addition, the grain growth and settling have been detected in disks (van Boekel et al. 2004; Furlan et al. 2006). Therefore, material fractionation is potentially operative during evolutionary stages prior to the proto-Sun achieving its final mass.

To model the potential enrichment of the Sun we must account for the total oxygen isotopic anomaly measured in the Sun, which is uncertain ($\delta^{18}O$=+50/-50 or $\Delta^{17}O$=+20/-20 ‰), and the Solar C/O ratio (~0.5) (Allende Prieto et al. 2002). The C/O ratio can be affected by material fractionation because the ices contain an excess of oxygen relative to the carbon (i.e. when the ices form most of the carbon is in the gas in the form of CO). If the initial C/O ratio was identical to the current Solar value, there would be no material fractionation functioning until the proto-Sun achieved its final mass, and the Solar $\delta^{18}O$ would be the assumed initial ISM value of -50 ‰ for all $G_0$. However, as mentioned above, material fractionation possibly operates before the proto-Sun achieved its final mass. Therefore, the initial C/O ratio in the Solar parent cloud may have been higher than the current Solar value since as material fractionation occurs, the total Solar $\delta^{18}O$ will increase, but the Solar C/O ratio will decrease. Coupled with this effect, we have found an initial C/O ratio of 0.7 provides the measured highest Solar $\delta^{18}O$ (+50 ‰) when the solar C/O ratio is determined by material fractionation. This ratio is within known interstellar values (see Bergin et al. 1997b for a discussion of C/O ratios in molecular clouds). However, lower values of the C/O ratio could account for lower Solar $\delta^{18}O$ ratios. Nonetheless, if material fractionation is active in the inner nebula as posited by Cuzzi et al. (2004) then the current Solar ratio is likely a lower limit to the ratio of the parent cloud.

We also assume that inside the snow line, $H_2O$ is enhanced compared to CO by a factor (f), which is a measure of the fact that grains of a certain size can accrete more quickly than gas through the disk (Ciesla et al. 2006). In addition, the isotopic fractionation is preserved from ~100 AU, which is the innermost boundary of our molecular core collapse model and the outer radius of the solar nebula, to the snow line located at 5 AU. Fig. 9 shows the C/O ratio and $\delta^{18}O$ in the proto-Sun versus $M_f$, the fraction of the Solar mass affected by the material fraction. The value of $M_f$ varies according to the time at which we assume significant grain growth in the midplane has occurred and material fractionation can begin. For example, if we assume that material fractionation begins at $4\times10^5$ year after the collapse, then only 0.1 $M_\odot$ out of total 1 $M_\odot$ ($M_f$=0.1) of the forming proto-Sun is affected.

Here, $\delta^{18}O_{SMOW}$ of the proto-Sun has been calculated based on the equation,

$$\delta^{18}O_{SMOW}\ (Protosun) = \frac{\int_0^{t'} \delta^{18}O_{snow\_line}\frac{dM}{dt}dt}{M(t')} \quad (1),$$



where $\delta^{18}O_{snow\_line}$ is the isotope anomaly of material crossing the snow line at a given time, $dM/dt$ is the mass of the accreting material at the time, and $t'=4.6\times10^5$ years, when the proto-solar mass becomes 1 $M_\odot$. $\delta^{18}O_{snow\_line}$ is given by the equation,

$$\delta^{18}O_{snow\_line} = \delta^{18}O_{grain}(Cloud)\{\frac{f_{grain}[O]_{grain}}{f_{grain}[O]_{grain} + f_{gas}[O]_{gas}}\} + \delta^{18}O_{gas}(Cloud)\{\frac{f_{gas}[O]_{gas}}{f_{grain}[O]_{grain} + f_{gas}[O]_{gas}}\}$$

(2), where $\delta^{18}O_{grain}(Cloud)$ and $\delta^{18}O_{gas}(Cloud)$ are isotope anomalies calculated from grains and gas, respectively, at 125 AU of the proto-solar cloud. Most of the oxygen in grains is in the form of water ice, with CO as the main reservoir of gaseous oxygen. $[O]_{grain}$ and $[O]_{gas}$ are oxygen abundances in grain and gas, respectively, and $f_{grain} = \frac{f}{f+1}$ and $f_{gas} = \frac{1}{f+1}$, where $f$ is the material fractionation between grain and gas. In this calculation, we use $f=1$ (no material fractionation) until the time when material fractionation begins (i.e. it takes some time for grains to grow to a size such that they are subject to radial gas drag forces).

In our calculations we must also account for the carbon and oxygen locked in the grain refractory component. Here we assume that carbon is in the form of graphite and oxygen is locked in silicates (FeMgSiO$_4$ as assumed in Lyons and Young 2005). These additional atoms are included in the calculation of the Solar C/O ratio and $\delta^{18}O_{SMOW}$ affected by the material fractionation. FeMgSiO$_4$ grains are not affected by photodissociation, and hence, have no oxygen isotope anomalies. For the initial abundance of carbon locked in grains we assume that the Solar carbon abundance arises from contributions from CO in the gas and carbon in grains, which corresponds to grains having an abundance that is 20 % of CO. The amount of oxygen locked in silicates can be calculated from our initial C/O ratio (0.7) and accounting for the various other oxygen reservoirs (H$_2$O ice and CO gas).

Based on Fig. 9(a), if 7-13 % of the Solar mass is affected by the material fractionation, then the Solar C/O ratio becomes 0.5, and the proto-Sun can have various $\delta^{18}O$ depending on $G_0$. As seen in Fig. 9(b), the Solar $\delta^{18}O=+50$ ($\Delta^{17}O=+25$) ‰ can only be accounted for with $G_0=25$ and $M_f=0.13$ while $-50$ ($\Delta^{17}O=-20$) ‰ can be matched by models with either $M_f=0.13$ ($G_0=1$) or 0.07 ($G_0=10^2$, $10^3$, and $10^5$). $G_0=10$ and $10^4$ can provide the Solar $\delta^{18}O$ of 0 ‰ and -30 ‰, respectively, at $M_f=0.13$ and $M_f=0.07$. If 10 % of the Solar mass is affected by the material fractionation, that is, $M_f=0.1$, material around 2,000 AU before collapse mainly contributes to the fractionation. As seen in this calculation, various values of Solar $\delta^{18}O$ are possible depending on $G_0$. *Therefore, accurate measurements of $\delta^{18}O$ from the Sun can constrain the radiation environment in which the Sun started to form.*

**V. Discussion**

The solar nebula beyond the snow line becomes deficient in water because of the material fractionation process. As a result, the total $\delta^{18}O$ in the outer solar nebula is almost the same as that of CO gas, which is very deficient in heavier isotopes. This is in contrast to the enhancements seen in meteorites and planetary bodies. Therefore, at later times, the water ices from the cloud to the outer disk should be highly enriched in heavier isotopes to compensate for its deficiency in the solar nebula. In our models we have 5 cases that can reproduce the solar $\delta^{18}O$ of either -50 ‰ or +50 ‰. Of these the $G_0=1$, $10^2$, and $10^3$ models do not provide water ices highly enriched in heavier isotopes to the solar nebula during the final accretion phase. This is due to the fact that anomaly peaks (Fig. 4) are located at too large radii to affect the anomalies in the solar nebula within a reasonable timescale. In the case of $G_0=25$, the enrichment is very large through out the cloud, so this is not an issue during any stage of collapse. Due to the strong radiation field (maximum anomalies are developed at small radii and move inward) for $G_0=10^5$, large enhancements can be provided to the solar nebula during the final stages of collapse.

In Fig. 10, to demonstrate this effect, we present the isotopic anomalies with time at the outer edge of the disk in the model of $G_0=1$, 25, and $10^5$ ($G_0=10^2$ and $10^3$ show similar values (< 100 ‰) to $G_0=1$). The time steps in this plot begins at $4.5\times10^5$ years, when the mass of proto-Sun is ~1 $M_\odot$, in the step of 5,000 years. $\delta^{17}O/\delta^{18}O$ follows the slope observed for CAIs. The CAIs follows the mass-independent fractionation line (MIF) as opposed to the mass-dependent terrestrial fractionation line (TFL). As seen, $\delta^{18}O$ in the outer edge of the solar nebula grows with time because higher anomalies imprinted in outer parts of the collapsing cloud are carried inward. In addition, unlike the case of $G_0=1$ (as well as $G_0=10^2$ and $10^3$), oxygen isotopic anomalies in the models of $G_0=25$ and $10^5$ are big enough to compensate for the deficiency of heavier oxygen isotopes developed by material fractionation in the outer solar nebula and thus to explain the observations in the Solar System. However, the $\delta^{18}O$ from comets (Eberhardt et al. 1995) are not much greater than 100 ‰. Water ice in comets is assumed not to be altered from the water ice provided into the solar nebula from the proto-solar cloud. Therefore, the model of $G_0=25$ provides anomalies that are too high if the measurements from comet Halley are representative of pristine comets. However, given the evidence for radial mixing in the nebula observed in Stardust samples (e.g., Brownlee et al. 2006), it is unlikely that cometary water can be compared isotopically to predicted pristine water. Also for $G_0=10^5$, accretion from the parent core to the disk should cease quickly after the proto-Sun has gathered its final mass since anomalies grow with time and reach the value obtained in $G_0=25$ at later time steps.

The $\delta^{17}O$ and $\delta^{18}O$ values for $H_2O$ ice in Fig. 10 agree qualitatively with recent measurements of nanocrystal aggregates in Acfer matrix (Sakamoto et al. 2007). The nanocrystals are a poorly characterized phase of an Fe-Ni-O-S bearing mineral that probably formed by oxidation of Fe metal or FeS by water in the solar nebula or in a planetesimal; Sakamoto et al. (2007) argue against formation on the Acfer parent body. δ–values for inferred $H_2O$ are ~ 150 – 200 ‰, and lie close to the CCAM and slope-1 lines. These are, by far, the most $^{16}O$-depleted solar system materials known. Large positive δ-values for $H_2O$ are predicted by all self-shielding models (Yurimoto and Kuramoto 2004; Lyons and Young 2005), but the largest values are predicted by the present model (Fig. 10) at the inner edge of the collapsing cloud. The values in Fig. 10 would most likely be significantly diluted at $\sim 3$ AU. The mass fraction of the nanocrystal aggregates is $\sim 100$ ppm in Acfer matrix, and the size of the implied $^{16}O$-depleted $H_2O$ reservoir is small. Whether the Acfer data is a residual fraction of a much larger $^{16}O$-depleted water reservoir, as is needed by the CO self-shielding models (Clayton 2002; Yurimoto and Kuramoto 2004; Lyons and Young 2005), remains to be seen.

Our study shows that various Solar $\delta^{18}O$ and $\delta^{17}O$ are possible depending on $G_0$, which can constrain the environment for the formation of the Sun. The best model constrained by measurements from meteorites, comet Halley, and the nanocrystal aggregates in Acfer matrix (Sakamoto et al. 2007) may be $G_0 = 10^5$ as seen in Fig. 10. However, this is clearly not a unique solution.

The inferred presence of short-lived radionuclides in meteorites (Wadhwa et al. 2007, and references therein) suggests the Sun formed near a massive star, which is consistent with $G_0 = 10^5$. The OB stars in a cluster of $\sim 4000$ stars (e.g., Trapezium in Orion) provide $G_0$ of about $10^5$ at a distance of $\sim 0.1$ to $0.3$ pc from the center of the cluster (Adams et al. 2004; Ouellette et al. 2005). Our results are consistent with the presence of a massive star formed coeval with the Sun in a large cluster, and in this model the bulk Sun would have $\delta^{18,17}O = -50$ ($\Delta^{17}O = -25$) ‰. Coeval, in this context, does not imply that the Sun and the massive star were of exactly equal age. It merely means that they overlapped for some period of time, before the massive star went supernova. Recent work by Bizzarro et al. (2007) shows evidence for the formation of differentiated planetesimals in the absence of $^{60}Fe$. Differentiated planetesimals are believed to have formed within $\sim 1$ Myr of solar system formation. If the massive star were 40 $M_{solar}$ with a lifetime $\sim 5$ Myr, the star would have existed for $\sim 4$ Myr before the birth of the solar system. Several authors have independently argued for this value of the Solar ratio (Clayton 2002; Yurimoto and Kuramoto 2004; Lyons and Young 2005), and our results also suggest that to obtain this ratio the presence of a massive star.

However, this model can also account for a range of possible Solar isotopic ratios. For instance, our model can also explain an enhancement in Solar $^{17}O$ and $^{18}O$ ($\delta^{18}O$, $\delta^{17}O \approx +50$ or $\Delta^{17}O = +20$ ‰) (Ireland et al. 2006) by invoking a moderate FUV radiation field ($G_0 = 25$) during pre-collapse. To dissociate CO this FUV radiation must be generated between $\sim 912 - 1000$ Å where the CO pre-dissociation bands lie. To generate sufficient FUV radiation requires the presence of at least a $\sim 5$ $M_{\odot}$ star (spectral class $> B8$) and more likely higher masses (see Fig. 1 in Parravano et al. 2003). Massive OB stars are only born in associations or stellar clusters (Blaauw 1992) and therefore even this weaker radiation field implies birth in a group of stars. An enhancement $G_0 = 25$ could arise by a nearby (0.1-0.5 pc) B star or a more massive O star with significant absorption between the proto-solar core and the massive star. The astrophysical context readily allows a situation where massive stars are exposed while other stars are still forming buried within absorbing molecular cloud material (e.g. the Trapezium and the Orion Molecular Cloud). In this case for an O star, which depending on its relative location can provide enhancement factors of $G_0=10^5$, would only need ~3 mag of dust absorption to reduce its radiation field on a forming molecular core to $G_0 \sim 25$. In this instance the implantation of $^{60}Fe$ directly into the solar nebular disk would be inhibited and the $^{60}Fe$ would need to be deposited into the molecular cloud that collapsed to form a centrally concentrated core and then onto the solar nebula.

To summarize we have presented a model that details the creation of oxygen isotopic anomalies on the surface of the proto-solar core by photo-chemical self-shielding of CO and its isotopologues, prior and commensurate with star birth. These anomalies are then delivered to the forming planetary disk via infall and advect inwards to feed the inner nebula where meteorites are born. Depending on the degree of decoupling during inward migration of gas and ice-coated dust grains (which carry isotopically enriched water ice), CO self-shielding can also alter the oxygen isotopic composition of the forming proto-Sun. Only proto-solar core models of isotopic enrichment via photo-chemical self-shielding predict altered solar isotopic composition, but all self-shielding models predict that the inner nebula should be enhanced in any isotopic (e.g. oxygen, nitrogen, etc.) enrichment potentially carried by dust grains. The external environment plays a key role as the presence of massive stars in the near vicinity can significantly alter the amount of isotopic enrichment, and thus the oxygen isotopic ratio of the Sun can constrain the original environment of the solar system. We find that our models suggest that the oxygen isotopic ratio of the Sun (based on two disparate measurements) require that the Sun is born in a stellar cluster in the presence of a massive star. The results from Genesis will resolve the current confusion regarding the Solar oxygen isotopic ratio, and with this model we can set constraints on the external environment in which the Sun was born.

## Acknowledgements

Support for this work was provided by NASA through Hubble Fellowship grant HST-HF-01187 awarded by the Space Telescope Science Institute, which is operated by the Association of Universities for Research in Astronomy, Inc., for NASA, under contract NAS 5-26555 and by the NSF under Grant No. 0335207. This work was also supported by the Korea Science and Engineering Foundation (KOSEF) under a cooperative agreement with the Astrophysical Research Canter for the Structure and Evolution of the Cosmos (ARCSEC). We are very grateful to Mike Jura for helpful discussions and comments. J.-E. Lee also thanks Ed Young and John Wasson for helpful discussions. J. R. Lyons gratefully acknowledges NASA Origins grant NNG06GD99G for support.

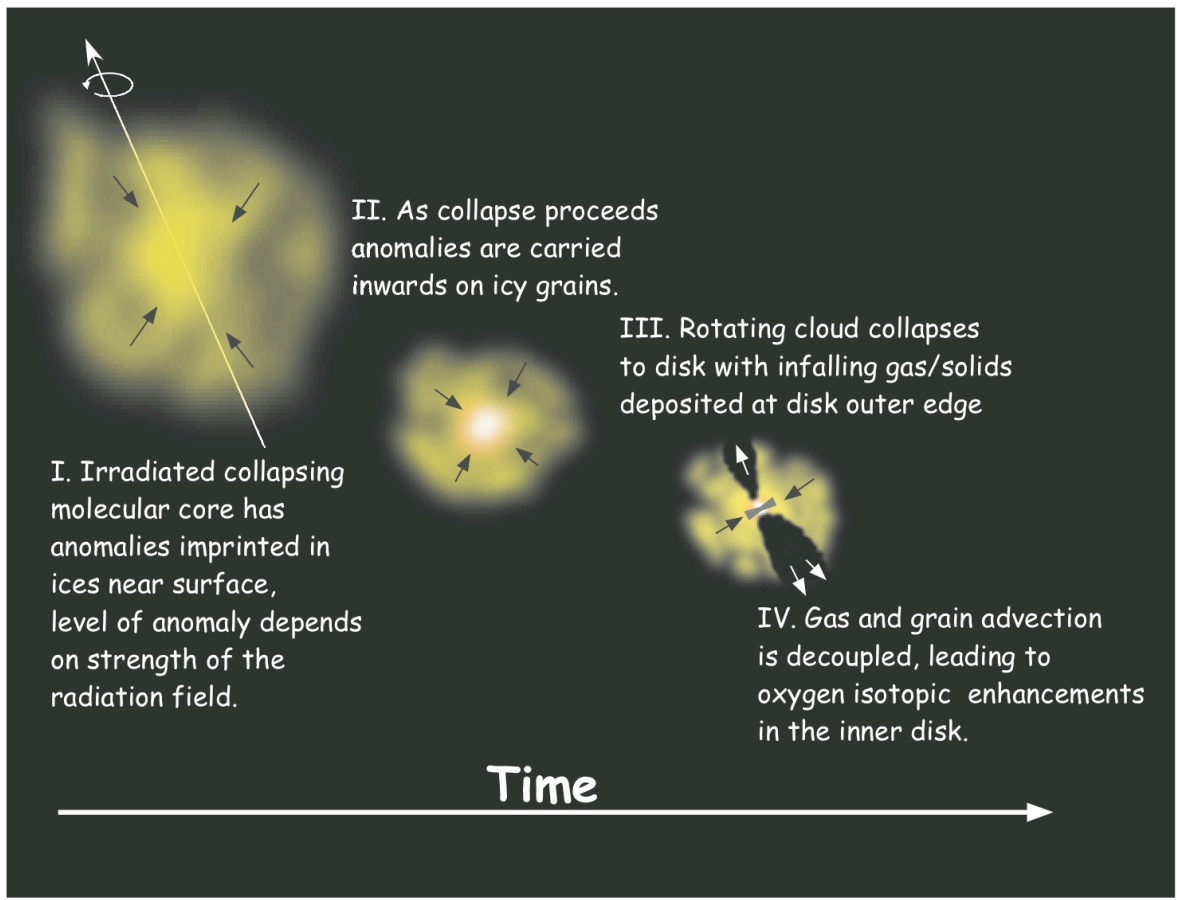

Fig. 1. The development of oxygen isotopic anomalies in a collapsing molecular core and the transfer of the anomalies to the disk, through which accretion occurs.

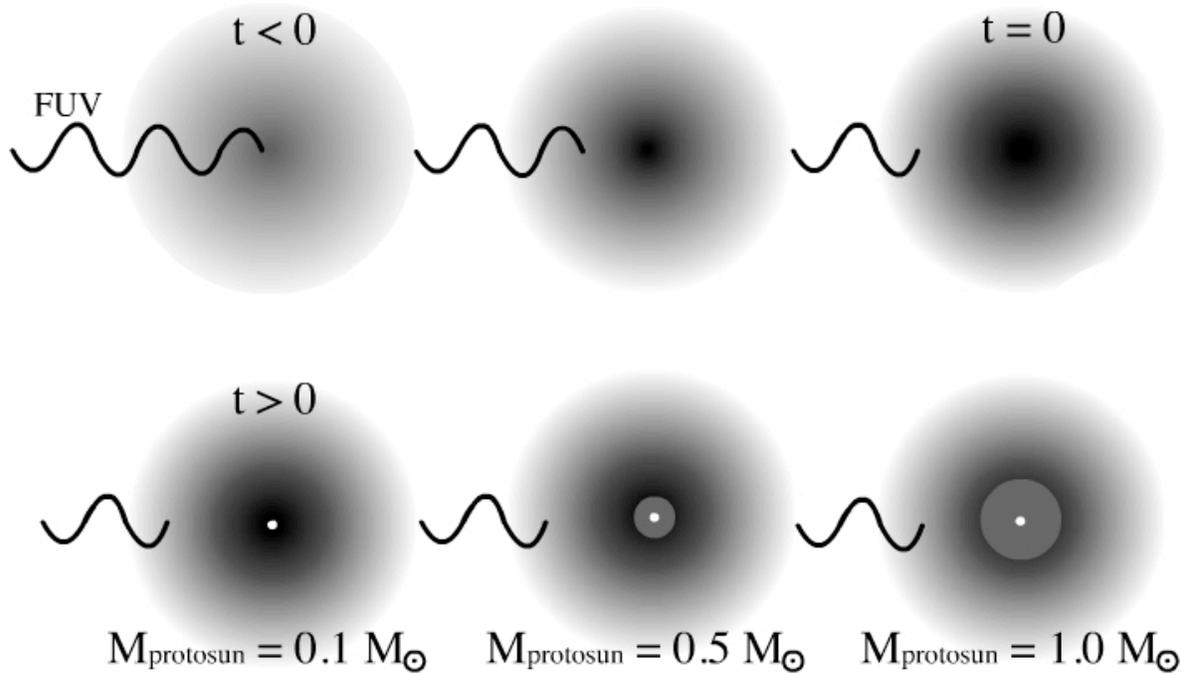

**Fig. 2**. A schematic diagram of the evolution of a parent Solar molecular core from its very early time step before collapse to later time steps after collapse. The central density of the core, modelled as a Bonnor-Ebert sphere, grows with time (darker shading) making it difficult for FUV photons to penetrate to the deep center. Once a protostellar object forms at the center by gravitational collapse, the density at the center decreases, and the infall radius, inside which material is falling into the center, propagates outward with time (inside-out collapse). In our model, the timescales of $M_{protosun}$=0.5 and 1 $M_\odot$ are $2.2 \times 10^5$ and $4.6 \times 10^5$ years, respectively.

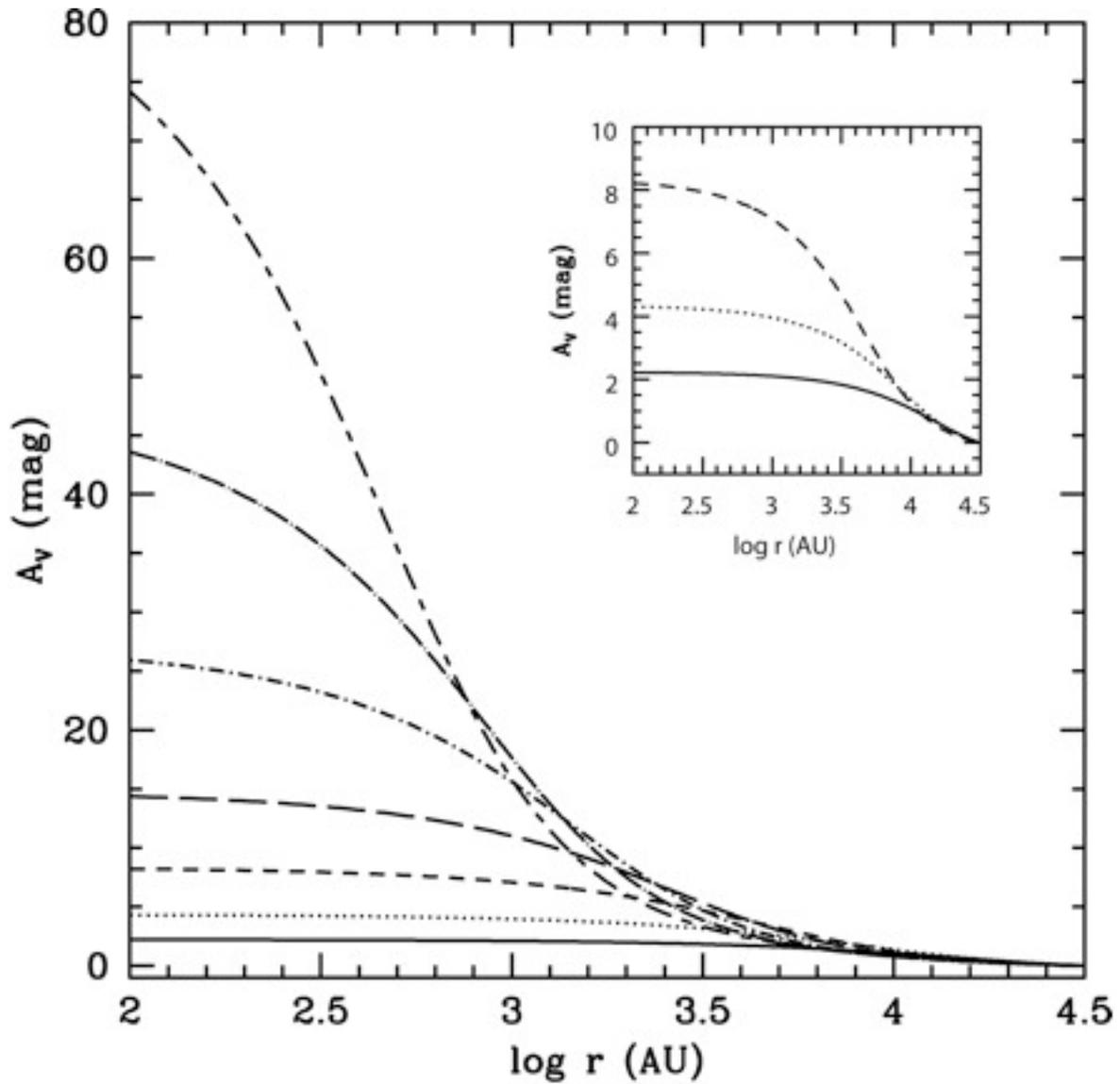

**Fig. 3.** The distribution of visual extinction ($A_V$) for 7 Bonnor-Ebert spheres, which describe the evolution of density profiles before collapse. Time steps of lines are $5 \times 10^5$ (solid line), $2.5 \times 10^5$ (dotted line), $1.25 \times 10^5$ (short dashed line), $6.5 \times 10^4$ (long dashed line), $3.5 \times 10^4$ (dot-short dashed line), $1.5 \times 10^4$ (dot-long dashed line), and $5 \times 10^3$ (short dash-long dashed line) years before collapse. In the upper right corner, three earliest time steps, which are the most important stages for the coupling between self-shielding of CO isotopes and shielding of FUV photons by dust grains, are brought up.

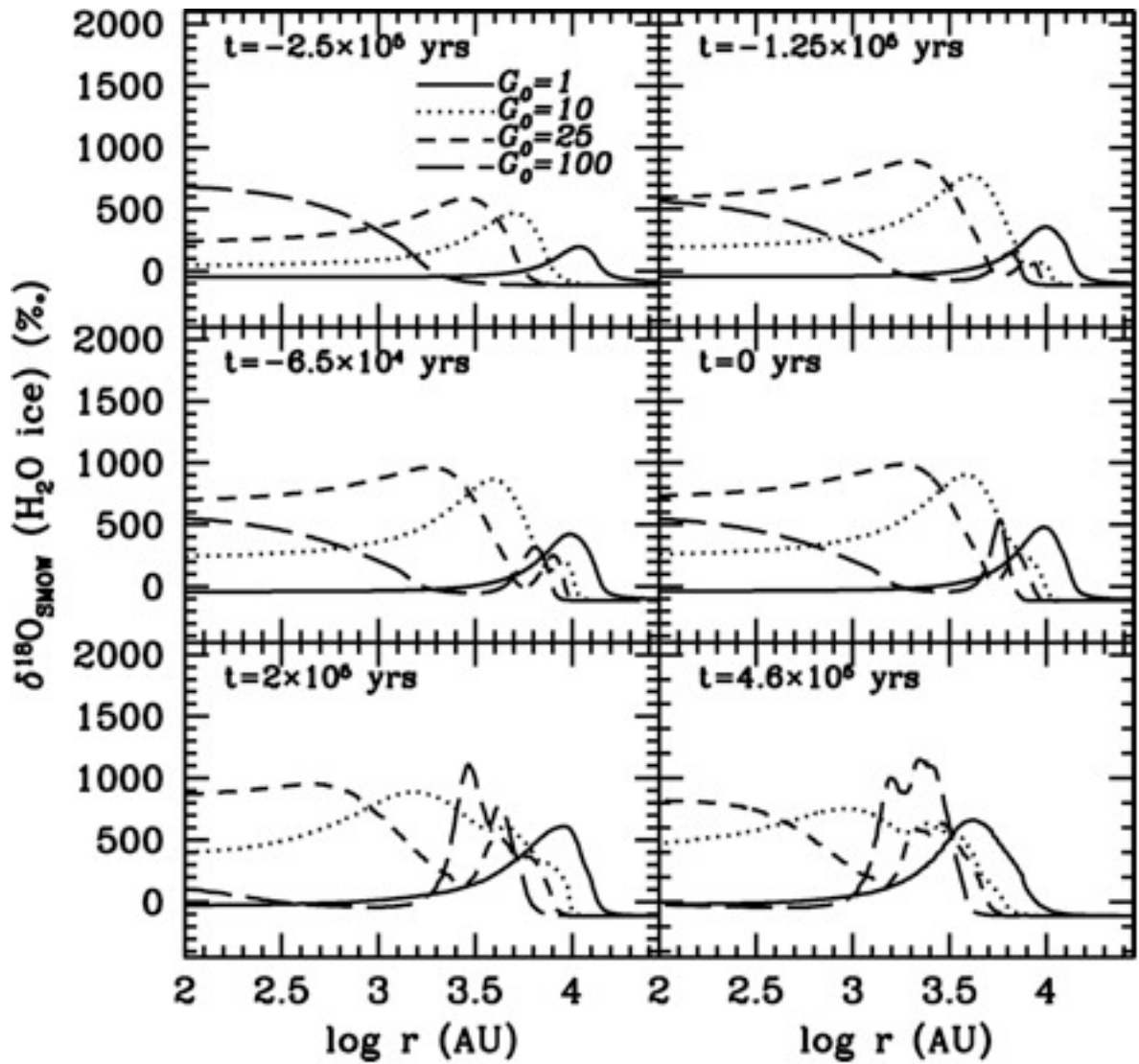

**Fig. 4.** Comparison of models with $G_0$=1, 10, 25, and $10^2$ at four time steps before and at 2× $10^5$ and 4.6 × $10^5$ yrs after collapse. At 4.6 × $10^5$ yrs, the total mass accreted to the protosun is about 1 M$_\odot$. $G_0$ is constant through the whole evolution in all models. $\delta^{18}O_{SMOW}$ is calculated for water ice. The coupled effect of CO shelf-shielding and dust self-shielding causes various levels of anomalies at the core center before collapse. After collapse, the peaks of anomalies are shifted inward. At 4.6 × $10^5$ yrs after collapse, the material inside ~20,000 AU is infalling to the central protostar (while material outside 20,000 AU is static), thus anomalies established in the outer regions are carried inward with the material. The material at 100 AU at 4.6 × $10^5$ yrs was originally located at 2,000 AU before collapse. However, in the case of $G_0$=1, the anomaly peak is developed at too large a radius to provide water ice highly enriched in heaver isotopes to the outer boundary of the solar nebula at later time steps.

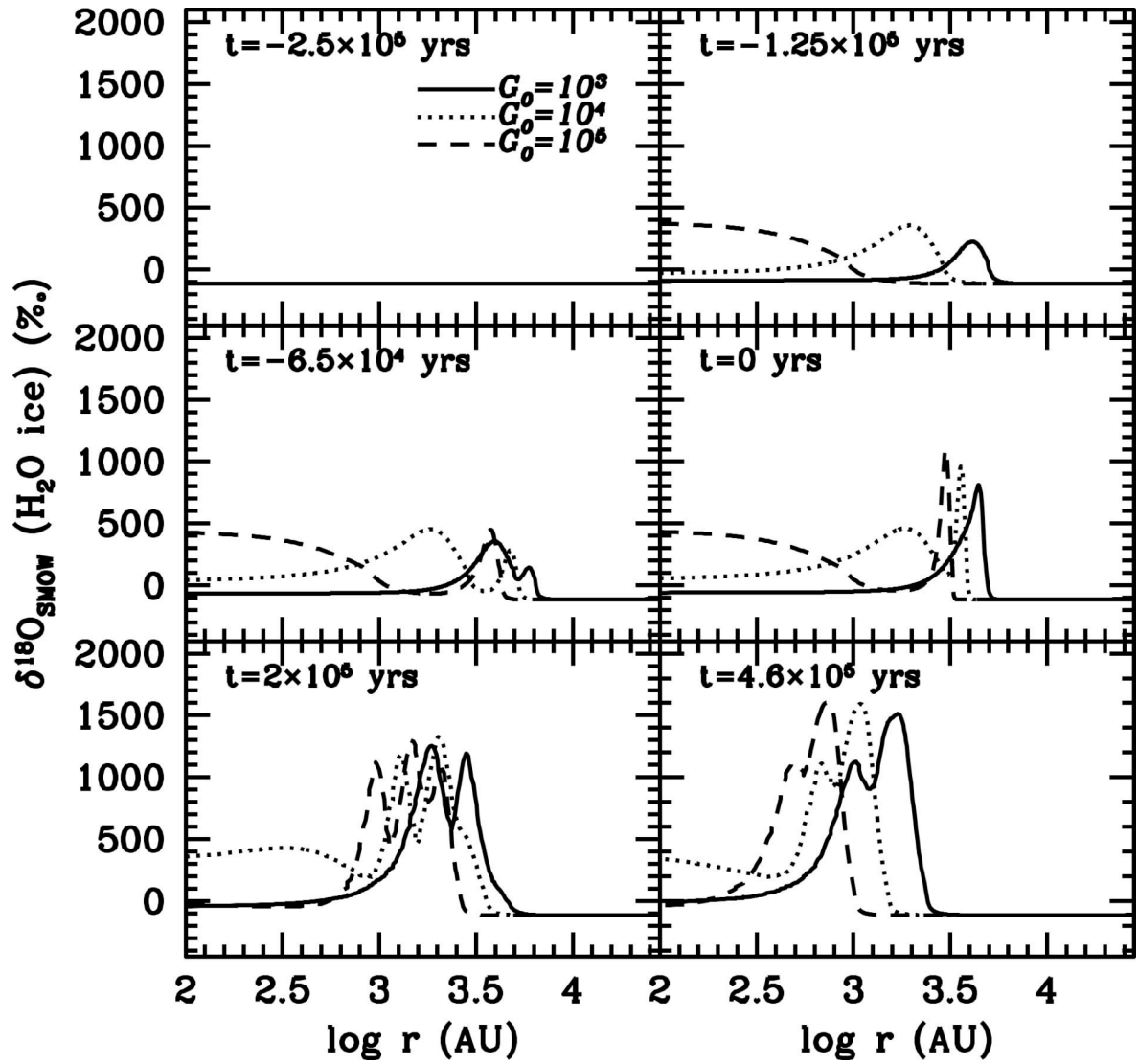

**Fig. 5.** Comparison of models with $G_0=10^3$, $10^4$, and $10^5$ at the same time steps as in Fig. 4.

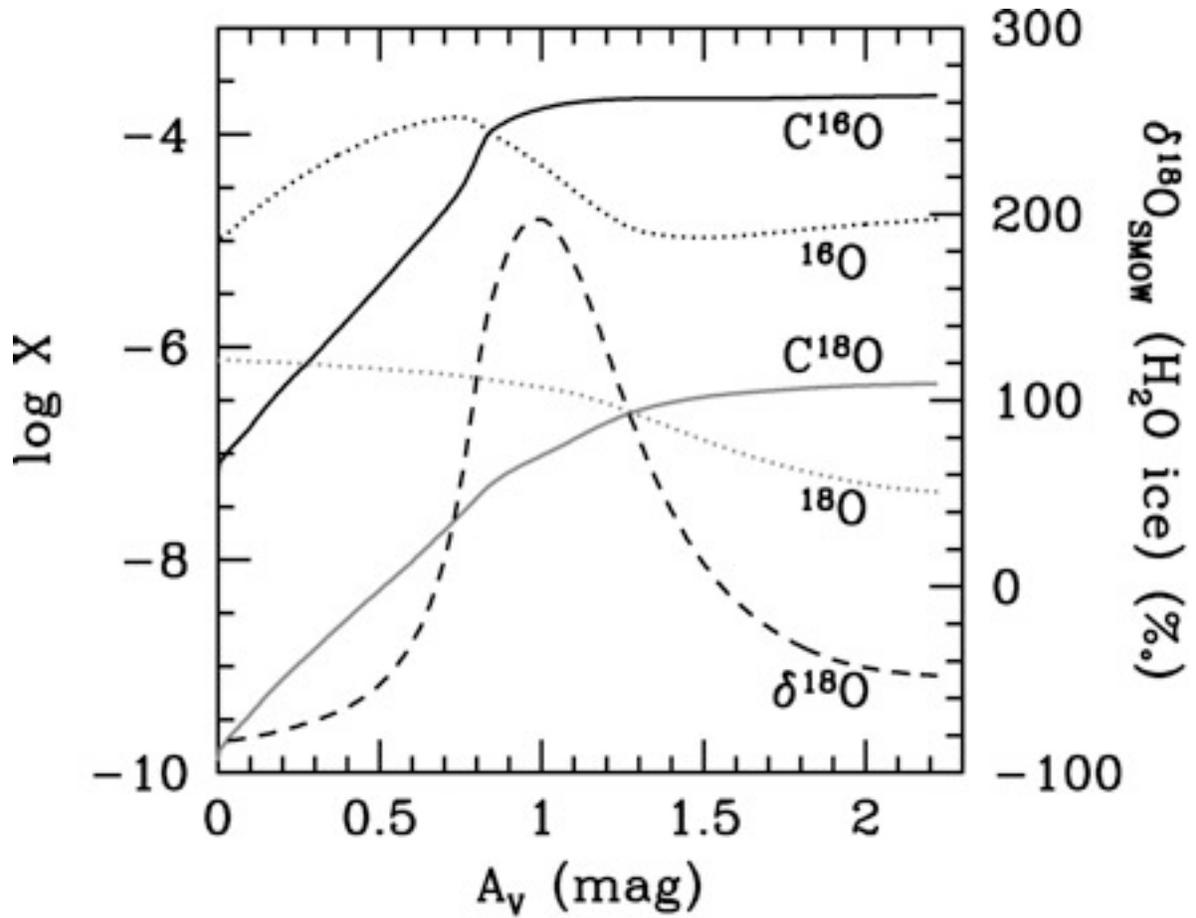

**Fig. 6** The distribution of C$^{16}$O, $^{16}$O, C$^{18}$O, and $^{18}$O abundances (left scale) and the oxygen isotopic anomaly, $\delta^{18}$O (right scale), at $-(2.5+\varepsilon) \times 10^5$ yrs before collapse with respect to A$_V$ for $G_0$=1. $\varepsilon$ is a very short timescale (~1 second here). A$_V$ is coupled to the photodissociation rates of CO isotopes, but is held constant during the calculation of chemical abundances at a given time step. The results shown here were computed with A$_V$ at a time step of -5 x 10$^5$ yrs (Fig. 3), which keeps constant until $-(2.5+\varepsilon) \times 10^5$ yrs. The anomaly peak is located at A$_V$=1 mag.

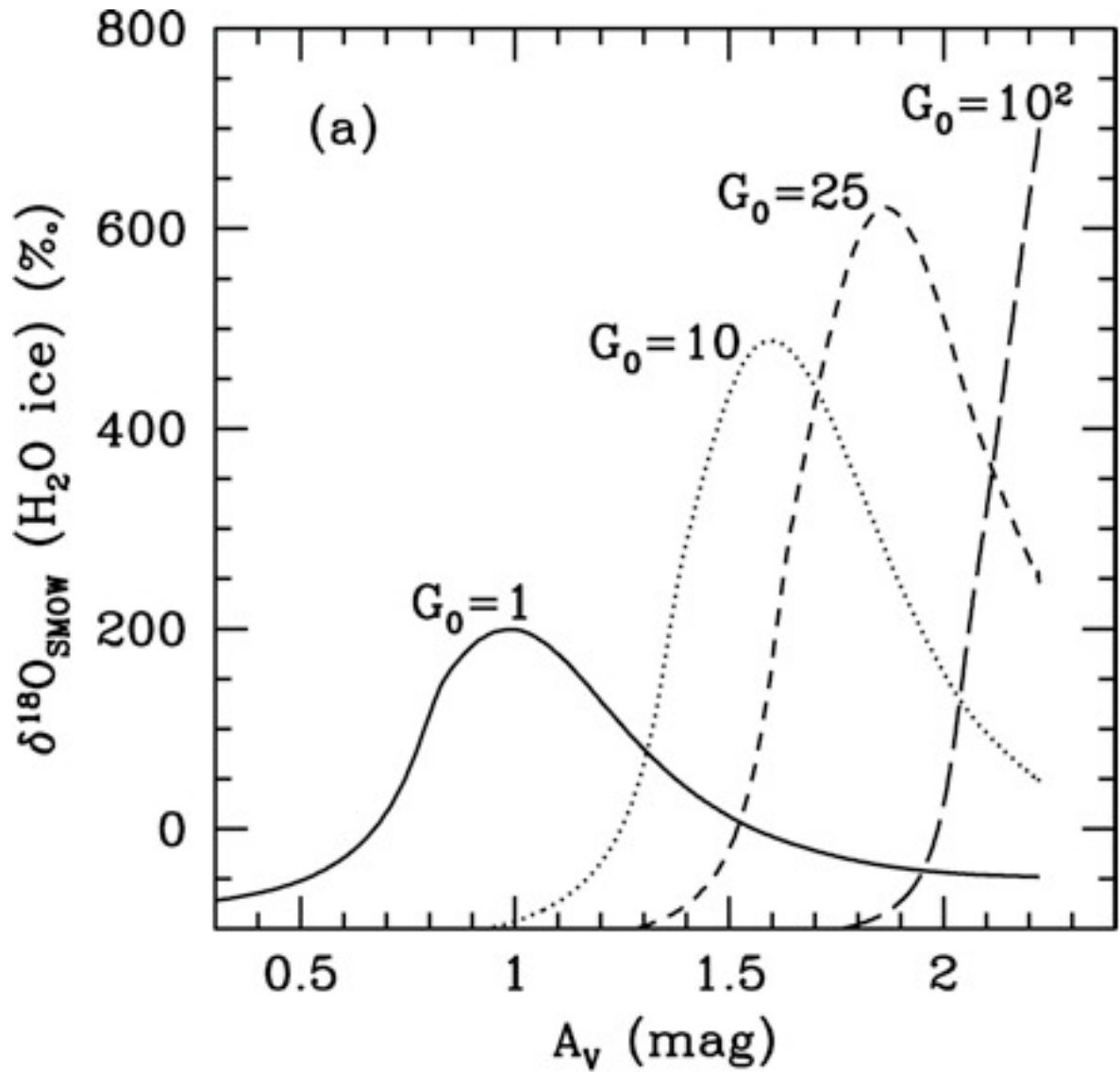

**Fig. 7 (a)** The distribution of the oxygen isotopic anomaly for various $G_0$. The time step for $\delta^{18}O_{SMOW}$ is the same as in Fig. 6. Anomaly peaks of $G_0$=10, 25, and $10^2$ are located at $A_V$=1.6, 1.85, and >2.2 mag.

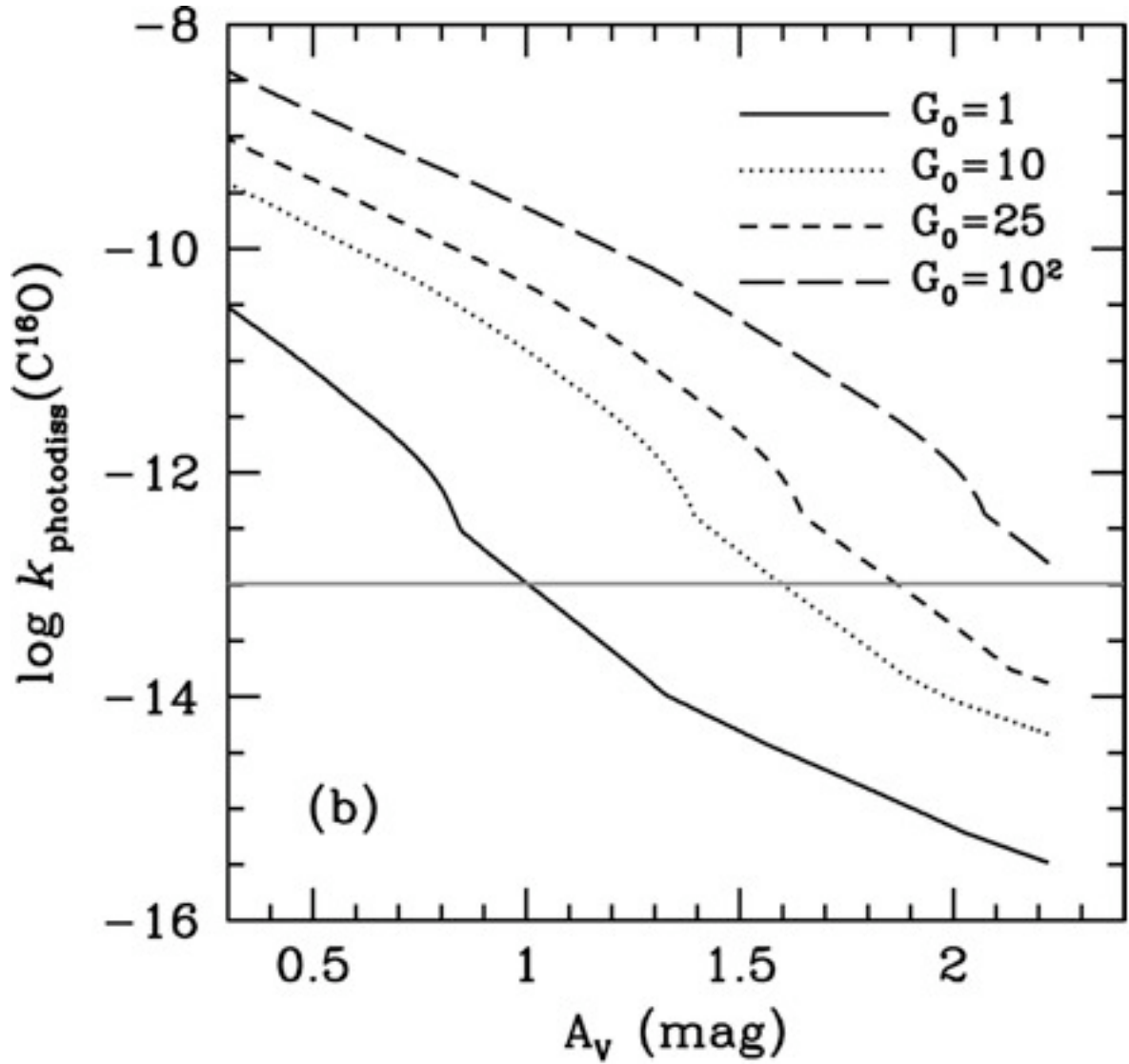

**Fig. 7 (b)** The distribution of the photodissociation rate ($k_{photodiss}$) of C$^{16}$O in various $G_0$ at $-(2.5+\varepsilon) \times 10^5$ yrs before collapse. The grey horizontal line indicates the photodissociation rate at the anomaly peak of $G_0=1$ ($k_{max}$), which is located at A$_V$=1 as seen in Fig. 6 and Fig. 7(a). The same photodissociation rate of C$^{16}$O occurs at A$_V$=1.6, 1.85, and >2.2 mag in $G_0$=10, 25, and 10$^2$, respectively. These A$_V$ are consistent with the A$_V$ where $\delta^{18}$O$_{SMOW}$ has peaks in $G_0$=10, 25, and 10$^2$ as seen in Fig. 7(a).

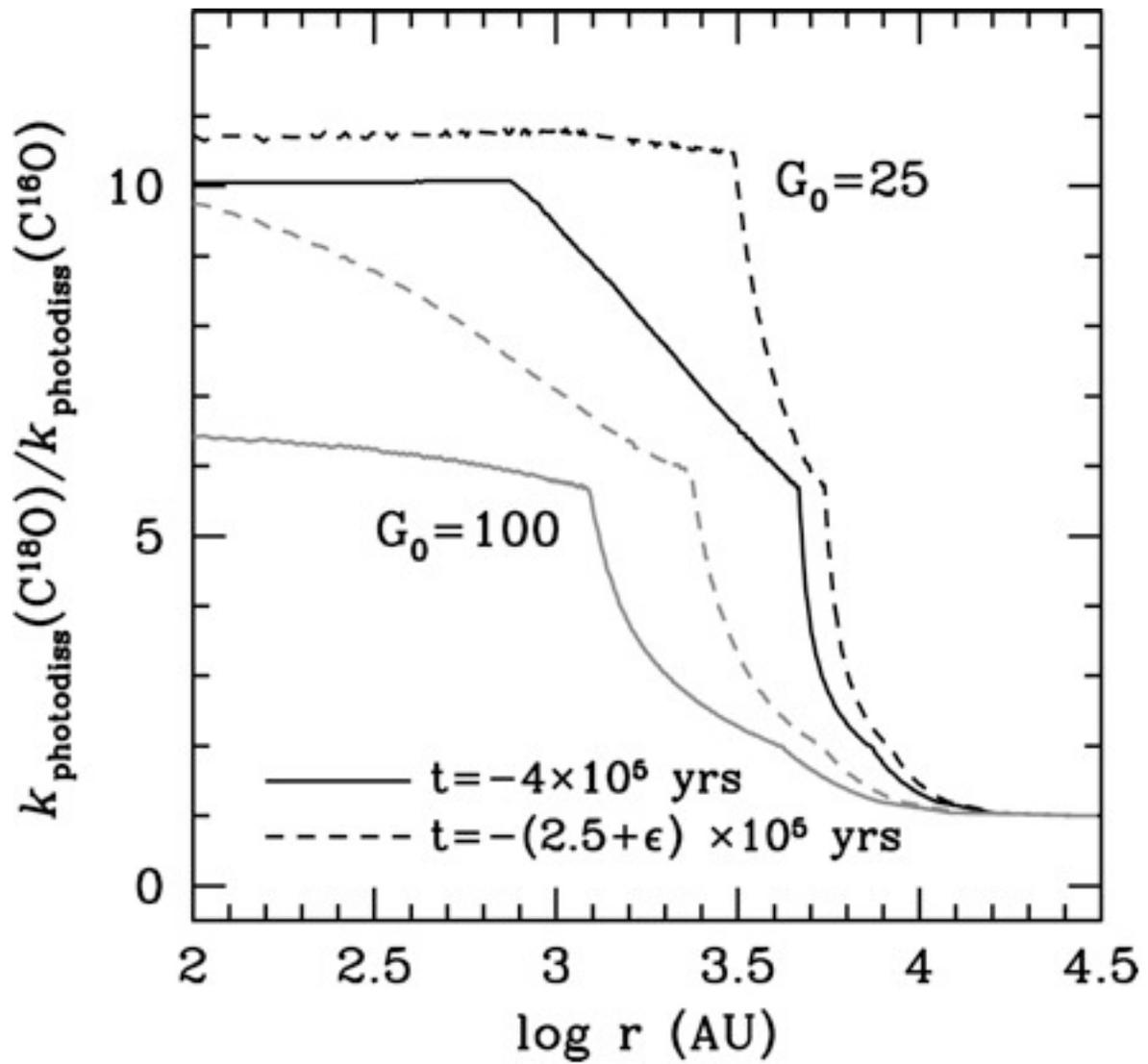

**Fig. 8** The distribution of the photodissociation rate ratio between $C^{18}O$ and $C^{16}O$ in $G_0 = 25$ (black lines) and 100 (grey lines) during the first time step of calculation (here seen at -4 × $10^5$ (solid line) and –(2.5+ε) × $10^5$ (dashed line) yrs before collapse).

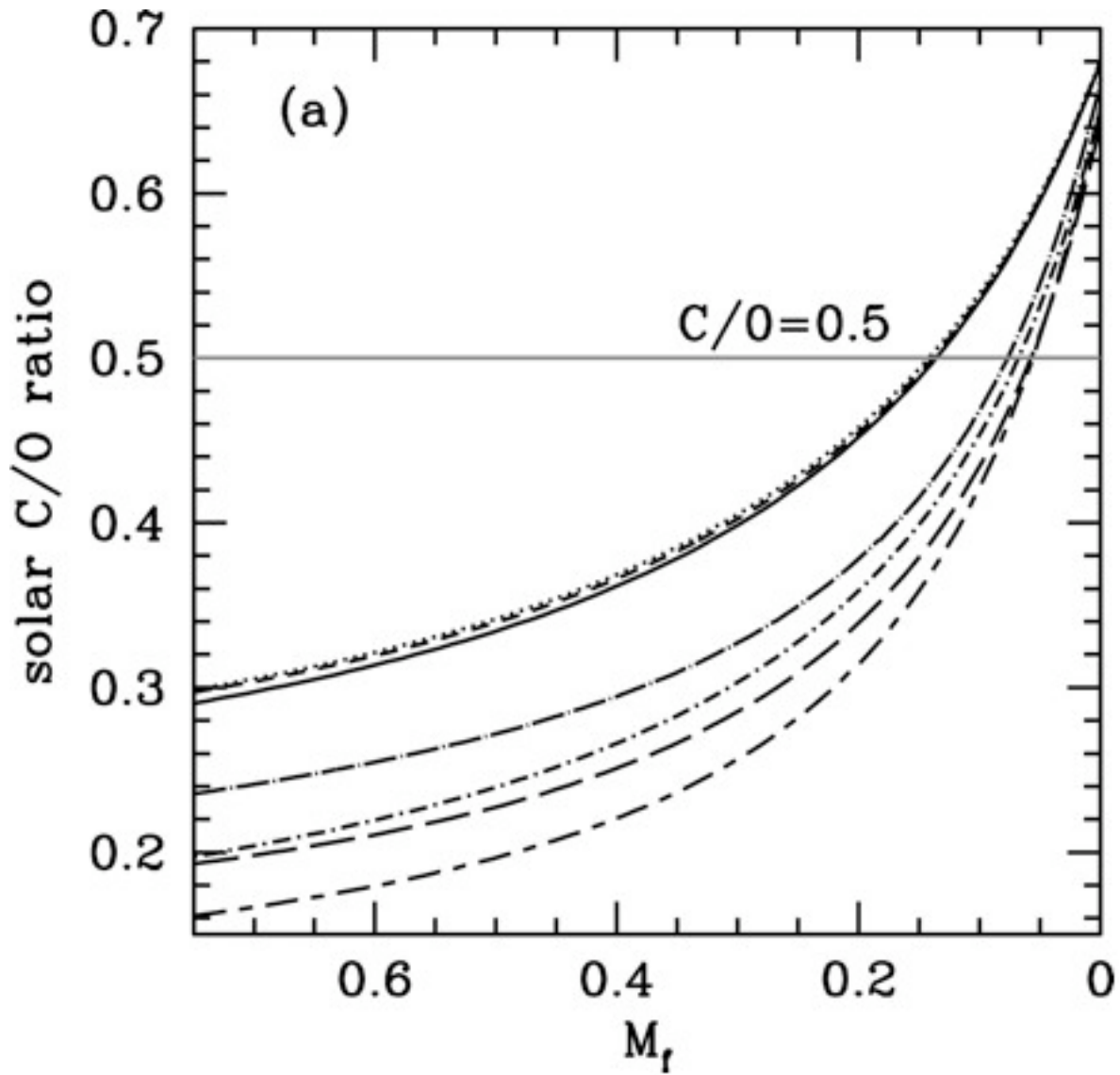

**Fig. 9 (a)** The variation of the C/O ratio in the proto-Sun with respect to the Solar mass affected by the material fractionation ($M_f$). In this calculation, we assume f=10, where f is the material fractionation between grains and gas, and the initial C/O ratio of 0.7. Horizontal grey line indicates the Solar value. This plot shows 7 (for $G_0 \gtrsim 10^2$) ~13 (for $G_0 \lesssim 25$) % of the Solar mass should be affected by material fractionation in order to provide the Solar C/O ratio (~0.5) when the protosun has 1 $M_\odot$. $M_f$=0.1 is equivalent with material fractionation that begins around t=4×10$^5$ years.

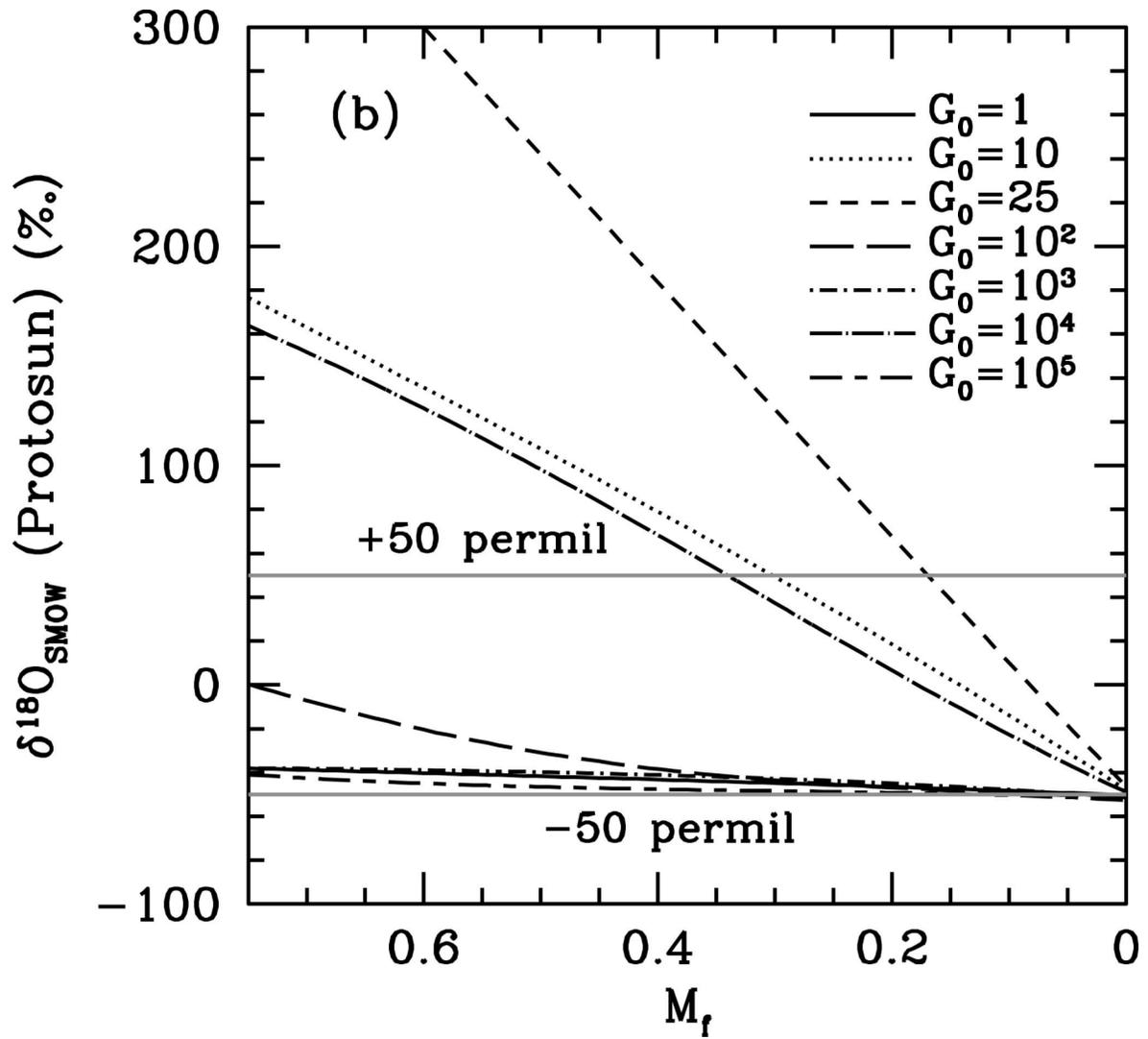

**Fig. 9 (b)** The variation of $\delta^{18}O_{SMOW}$ in the proto-Sun versus $M_f$. f=10 is assumed in this calculation. Various Solar $\delta^{18}O_{SMOW}$ values are possible depending on $G_0$ and $M_f$. At a given $G_0$, $M_f$ is constrained by the Solar C/O ratio (Fig. 10(a)).

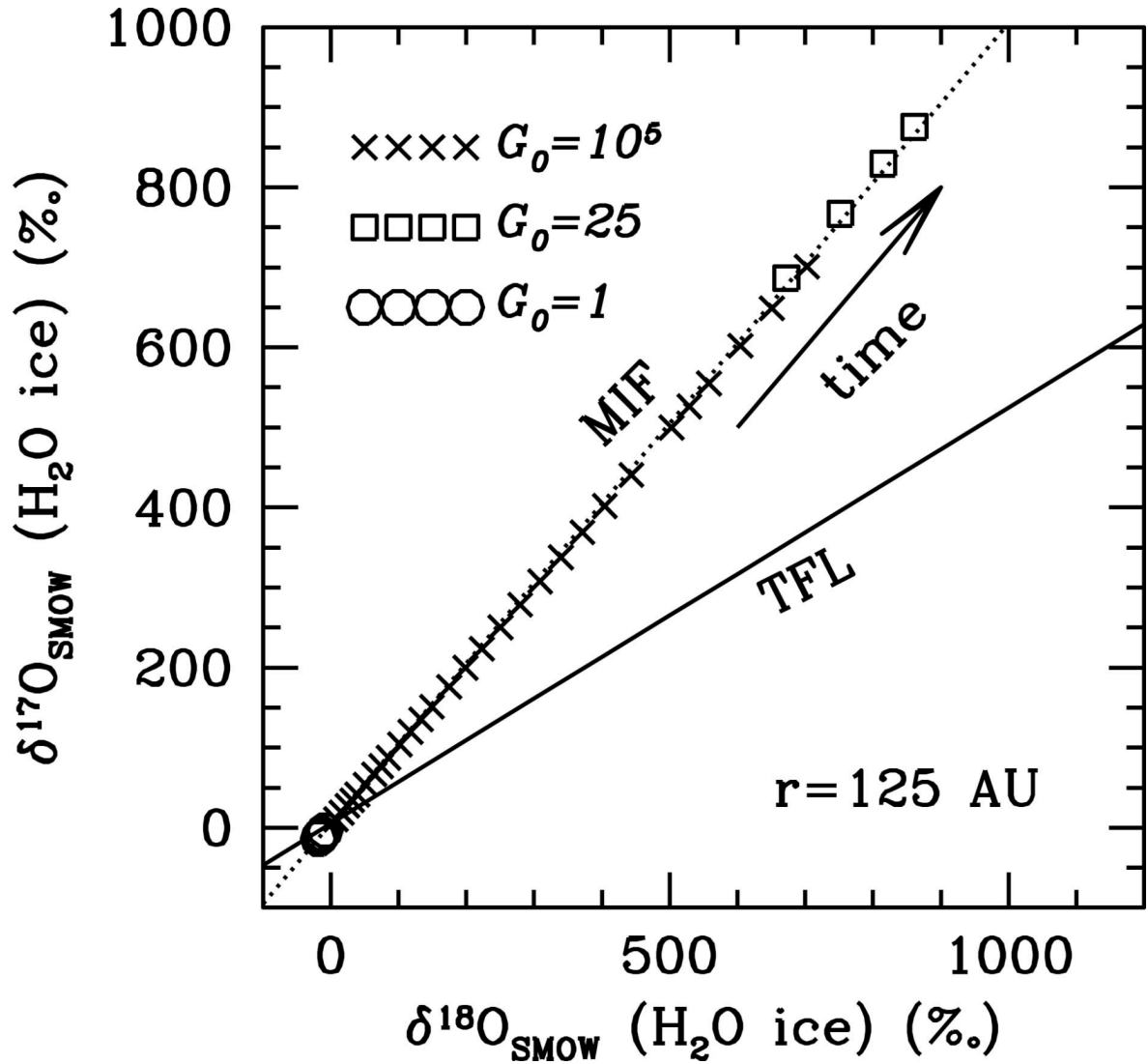

**Fig. 10.** The evolution of isotope anomalies calculated from water ices at the outer boundary of the solar nebula (r=125 AU) in the models of $G_0$=1 (circles), 25 (squares), and $10^5$ (x symbols). $\delta^{17}O_{SMOW}$ is calculated using the same equations for $\delta^{18}O_{SMOW}$ except the isotope ratio, $^{16}O/^{17}O$=2600. In the model of $G_0$=$10^5$, time steps for symbols starts at t=4.5×$10^5$ years and increase in the step of 5,000 years up to t=6×$10^5$ years. In the models of $G_0$=1 and 25, however, the time step for the plot is 50,000 years to avoid confusion. The isotope anomalies in these two models do not vary much with time. Anomalies calculated from water ices plot along the line with a slope of 1, the MIF (dotted line) rather than the TFL (solid line), as seen in meteorites.